\documentclass[11pt]{article}
\usepackage[margin=1in]{geometry} 
\usepackage{lipsum} 
\usepackage{graphicx} 
\usepackage{titlesec} 
\usepackage{natbib} 
\usepackage{authblk} 
\usepackage[T1, OT1]{fontenc}
\DeclareTextSymbolDefault{\dh}{T1}
\usepackage{hyperref} 
\setcitestyle{numbers,square} 

\usepackage[smartEllipses]{markdown}
\titleclass{\subsubsubsection}{straight}[\subsubsection]
\newcounter{subsubsubsection}[subsubsection]
\renewcommand\thesubsubsubsection{\thesubsubsection.\arabic{subsubsubsection}}
\titleformat{\subsubsubsection}{\normalfont\normalsize\itshape}{\thesubsubsubsection}{1em}{}
\titlespacing*{\subsubsubsection}{0pt}{3.25ex plus 1ex minus .2ex}{1.5ex plus .2ex}
\makeatletter
\renewcommand\paragraph{\@startsection{paragraph}{5}{\z@}%
{3.25ex \@plus1ex \@minus.2ex}%
{-1em}%
{\normalfont\normalsize\bfseries}}
\renewcommand\subparagraph{\@startsection{subparagraph}{6}{\parindent}%
{3.25ex \@plus1ex \@minus .2ex}%
{-1em}%
{\normalfont\normalsize\bfseries}}
\def\toclevel@subsubsubsection{4}
\def\toclevel@paragraph{5}
\def\toclevel@paragraph{6}
\def\l@subsubsubsection{\@dottedtocline{4}{7em}{4em}}
\def\l@paragraph{\@dottedtocline{5}{10em}{5em}}
\def\l@subparagraph{\@dottedtocline{6}{14em}{6em}}
\makeatother
\title{Revisiting gender bias research in bibliometrics:\\
Standardizing methodological variability using\\
Scholarly Data Analysis (SoDA) Cards}

\author[1]{HaeJin Lee}
\author[2]{Shubhanshu Mishra}
\author[1]{Apratim Mishra}
\author[1]{Zhiwen You}
\author[3]{Jinseok Kim}
\author[1,4]{Jana Diesner}

\affil[1]{School of Information Sciences, University of Illinois Urbana–Champaign}
\affil[2]{\href{https://shubhanshu.com}{shubhanshu.com}}
\affil[3]{School of Information, University of Michigan - Ann Arbor}
\affil[4]{School of Social Sciences and Technology, Technical University of Munich}

\begin{document}
\maketitle
\begin{abstract}

Gender biases in scholarly metrics remain a persistent concern, despite numerous bibliometric studies exploring their presence and absence across productivity, impact, acknowledgment, and self-citations. However, methodological inconsistencies, particularly in author name disambiguation and gender identification, limit the reliability and comparability of these studies, potentially perpetuating misperceptions and hindering effective interventions.  A review of 70 relevant publications over the past 12 years reveals a wide range of approaches, from name-based and manual searches to more algorithmic and gold-standard methods, with no clear consensus on best practices. This variability, compounded by challenges such as accurately disambiguating Asian names and managing unassigned gender labels, underscores the urgent need for standardized and robust methodologies. To address this critical gap, we propose the development and implementation of ``Scholarly Data Analysis (SoDA) Cards."  These cards will provide a structured framework for documenting and reporting key methodological choices in scholarly data analysis, including author name disambiguation and gender identification procedures. By promoting transparency and reproducibility, SoDA Cards will facilitate more accurate comparisons and aggregations of research findings, ultimately supporting evidence-informed policymaking and enabling the longitudinal tracking of analytical approaches in the study of gender and other social biases in academia.
\end{abstract}

\section{Introduction}
Gender inequality and bias in academia persist through systemic discrimination, implicit prejudices, and structural barriers, continuing to disadvantage women across various stages of their careers, hindering both their development and productivity \cite{belingheri2021twenty, casad2021gender}. These inequities and biases manifest in multiple forms, such as biased hiring evaluations and unequal access to funding \cite{carlsson2021gender,schmaling2023gender}. Such challenges create a compounded disadvantage, adversely affecting female researchers' recruitment, retention, promotion, and overall career progression \cite{heilman2001description, shen2013inequality, williams2015national}. Over time, these cumulative effects lead to the persistent underrepresentation of women at various stages in academic roles. For instance, according to the 2023 CRA Taulbee Survey \cite{cra_taulbee_2023}, women represented approximately 22\% of bachelor’s degree recipients, around 20\% of master’s degree recipients, and roughly 23\% of new Ph.D. graduates in Computer Science and closely related fields, consistently falling short of parity with their male counterparts across all degree levels. These disparities become even more pronounced in senior academic roles \cite{aauw_fast_facts}: women represent just 15\% of tenure-track engineering faculty and 14\% of tenure-track computer science faculty.

Such disparities are reported to be driven in part by ingrained cultural biases and ``masculine defaults'', underscoring the need to address systemic factors rather than viewing the problem solely at the individual level \cite{ridgeway2014status,cheryan2020masculine,kramer2023beyond}. Understanding and addressing these layers of gender inequalities and biases is crucial for fostering an inclusive and transformative intellectual landscape. To this end, numerous studies have examined the extent of gender bias in terms of scholarly metrics (e.g., number of citations and h-index) in academia \cite{merriman2021gender, salerno2019male, nunkoo2020three, murphy2020open,mishra_self-citation_2018,maggio2023voices, hagan2020women, fox2018patterns, ahmadia2021limited}. For instance, studies have continued to report that papers authored by male first authors receive significantly higher citation counts than those authored by female first authors \cite{thelwall2020gender,chatterjee_gender_2021}. Other studies have demonstrated both persistent gender disparities in scholarly recognition and authorship patterns: an analysis of cardiovascular research articles by \citet{chatterjee_gender_2021} found that women were underrepresented as both primary and senior authors: out of 5,322 primary authors, 35.6\% were women, and out of 4,940 senior authors, 25.8\% were women. Similarly, \citet{mohammad-2020-gender} observed that women accounted for 29.7\% of all authors, 29.2\% of first authors, and 25.5\% of last authors. Papers with female first authors received, on average, 37.6 citations, considerably fewer than the 50.4 citations garnered by papers with male first authors. Such patterns of under-representation and lower citation counts are consistent with additional research across various scientific disciplines \cite{odic_publication_2020}, underscoring the continued influence of gender bias on scholarly recognition. However, there are also studies that report the reduction, absence, or reversal of gender bias \cite{mishra_self-citation_2018, squazzoni2021peer,nielsen_gender_2016, huang_historical_2020} e.g. \citet{mishra_self-citation_2018} reported that gender bias in self-citation goes away when controlled for author's career length. These seemingly contradictory findings highlight the complex nature of gender bias in academia, with factors like career length often playing a crucial role. These insights can be leveraged to design and implement effective interventions that address the nuanced challenges faced by women in academia.

These systemic issues have the potential to affect policy decisions and perpetuate gender inequality and bias in academia while underscoring the need to identify, articulate, and address these biases. Accurately understanding the extent of gender bias is crucial, as these findings guide evidence-based policy decisions to mitigate this impact. It is, therefore, essential to ensure that the underlying research methods of gender bias studies are rigorously designed, executed, validated, and standardized to produce reliable and actionable results. However, existing studies often exhibit methodological inconsistencies, particularly in how authors’ names are disambiguated and genders are identified. Such variations in approach can significantly influence accuracy and reliability \cite{kim2016distortive,mishra_self-citation_2018,Wu2024GenderCitationGap}. These discrepancies within and between methodology pipelines may lead to unreliable conclusions about gender effects and differences in scholarly metrics and emphasize the importance of developing and adopting a more standardized approach to designing and conducting reliable studies. 

\begin{figure*}[htp]
\centering
\includegraphics[width=14cm]{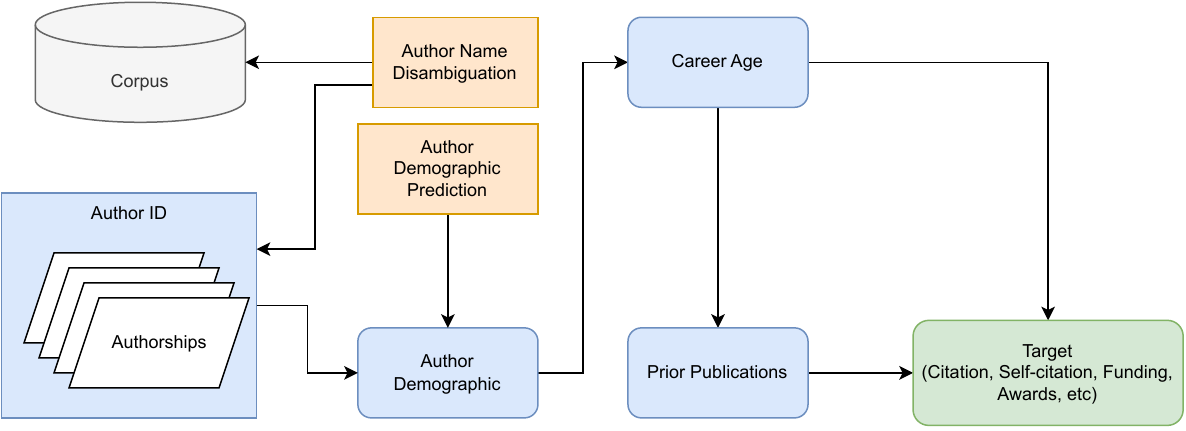}
\caption{A principled approach for conducting demographic bias analysis for scholarly data research}
\label{fig:flow_chart}
\end{figure*}

To address these challenges, we conducted a literature review of 70 papers published over the past 12 years, focusing on the methods used for author name disambiguation and gender identification. Through this review, we gained insights into recent practices and identified key inconsistencies. Building on these insights, we propose “Scholarly Data Analysis (SoDA) Cards,” a standardized framework for documenting and reporting methodological details in gender bias studies. By facilitating more accurate comparisons and aggregations across research, SoDA Cards can enhance evidence-based policy-making and help track methodological and analytical improvements over time. By developing SoDA Cards, we also identified a principled workflow (figure \ref{fig:flow_chart}) which we suggest for conducting future scholarly data analysis research. In the following sections, we explain the motivation and objectives of our study, describe the methodology of our systematic literature review and annotation process, present our findings, and introduce the SoDA Cards. 

\section{A landscape of scholarly data analysis research}

Scholarly data analysis is a broad research topic that focuses on the analysis of scholarly research publications. This ranges from analysis of citation networks of research papers and authors, to text analysis of research papers. Our work focuses on the practice of working with scholarly data with a focus on identifying author demographic-based patterns in scholarly metrics like citations, publication count, research careers, etc. To conduct such analysis, authors often utilize a few common methods like author name disambiguation and author demographic assignment, e.g., identification of gender, age, ethnicity, and race. Furthermore, scholarly data analysis often focuses on causal relationships between author demographics and scholarly metrics. We describe these methods and analysis in the following sections and highlight how the common methods in existing scholarly work motivated us to analyze the distribution of these methods in popular research and propose the construction of the SoDA Cards to standardize the practice of reporting scholarly data analysis practices in future works.

\subsection{Author name disambiguation}
\begin{figure*}[htp]
\centering
\includegraphics[width=\linewidth]{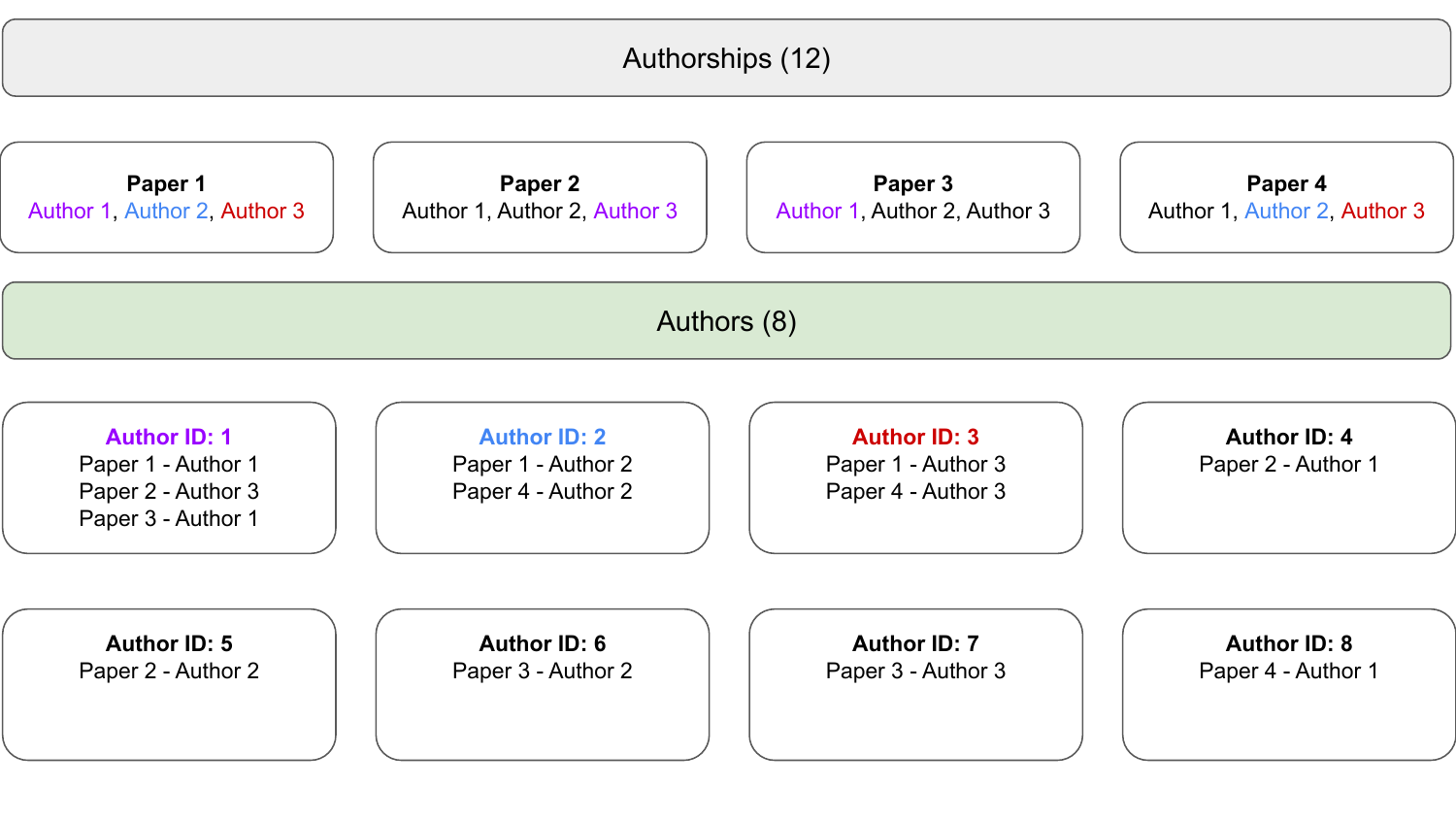}
\caption{Difference between author and authorship}
\label{fig:authorships}
\end{figure*}
Research papers list individuals involved with the research work in the paper byline. These individuals are often identified via their name, institution, and other identifiers. For the purpose of analysis, these listed identifiers on a research paper are called authorships, while the individuals associated with these identifiers are considered authors. This difference is illustrated in figure \ref{fig:authorships}. \citet{Wu2024GenderCitationGap} also provides a good overview of gender bias in citation research based on authorships and authors. The process of going from a list of authorships to authors is commonly referred to as author name disambiguation or AND.

Author name disambiguation is an essential task in bibliometric research. It involves accurately identifying and distinguishing unique authors, accounting for variations in names that may be identical or similar, and finding and indexing variations of unique authors, which is a frequent challenge in large bibliometric databases \cite{kim2014impact}. Typically, author names are disambiguated using similarity measures to assign authorship of bibliographic records to individual authors based on data from a compiled dictionary or bibliographic databases. AND is particularly important in gender bias studies, as it ensures the precise attribution of publications to the correct author. Ambiguities in author names can lead to misinformation about scholarly output, research impact, and patterns of collaboration \cite{kim2016distortive, mishra_self-citation_2018}. The goal of AND is to distinctly identify and index or represent each author, eliminating biases associated with names that could affect subsequent gender identification analyses. 

There are four primary approaches to author name disambiguation, each with its advantages and limitations. The first approach relies solely on heuristic or name-based methods, using partial or exact matches of names \cite{caplar_quantitative_2017,dworkin_extent_2020}. While this method is straightforward and easy to implement, it can be prone to errors, e.g., in cases where multiple authors share similar names. The second approach uses an algorithm that compare name strings and additional features such as affiliation to disambiguate authors' names \cite{liu_gender_2023,pinheiro_women_2022,mihaljevic2016effect}. The third approach involves manual searches, such as consulting online profiles, databases, or institutional directories \cite{maliniak_gender_2013,mayer2018does}. Although this strategy can yield highly accurate results, it is labor-intensive, time-consuming, and does not scale to large datasets. Fourth, some studies use self-reported data or draw upon identities established through a gold-standard process \cite{eloy2013gender,tao2017gender}. The gold-standard process typically involves curating a validated reference set of author identities by carefully verifying author information through trusted sources or expert judgment. This method provides a reliable benchmark for evaluating and refining other disambiguation techniques, but it demands extensive effort and resources to establish and maintain the reference set. 

Author name disambiguation is challenging for many reasons, such as variations in the spelling of names, the indexing of authors via initials instead of full names, inconsistent reporting of affiliation changes, and cultural differences in the ordering of names. Additional complexities arise, particularly with names of Asian origin as these names may not adhere to Western naming conventions, which increases the risk of misidentification of authors \cite{kim2016distortive, ijcai2024p0800}. Such misidentification can introduce significant biases into study results by incorrectly attributing publications to the wrong authors or failing to recognize distinct individuals as unique authors. For example, in the case of Asian-origin names, common surnames such as ``Kim" or ``Wang" combined with limited given-name differentiation can result in multiple authors being grouped under a single identity or a single author being split into multiple records \cite{kim2015effect}. Therefore, the effectiveness of the chosen author name disambiguation method in accurately distinguishing between authors is critical. Ensuring the precise identification of authors, particularly those with Asian-origin names, is essential to maintain the validity and reliability of gender bias studies in bibliometric research.

\subsection{Gender identification}
Once an author has been correctly identified, associating them with further demographic information often uses only the author's name. The task of identifying the gender of authors in bibliometric studies is another critical aspect in the investigation of gender bias. This data can then be used to assess patterns of gender bias within a research field. Traditional gender identification methods have relied on gender-name dictionaries or lists, which associate first names with a specific gender based on societal norms or legal documents \cite{lariviere_bibliometrics_2013}. These databases are often built from census data or similar large-scale demographic datasets. While effective for many Western names, these methods can struggle with unisex or non-Western names, leading to inaccuracies in gender assignment \cite{you-etal-2024-beyond}.

Recently, more sophisticated methods of gender identification have been developed to address these limitations. Some studies utilize machine learning algorithms trained on large datasets to predict the gender of an author based on their first name. These models can incorporate additional features, such as a name's cultural, institutional, or linguistic context, improving their accuracy, especially for non-Western names. Despite these advancements, gender identification in bibliometric studies still faces numerous challenges. For instance, handling unknown or ambiguous gender remains a significant issue. Many studies default to a binary gender model, which can inadvertently exclude or misclassify authors who do not identify as male or female. Furthermore, cultural and regional differences in names can complicate gender identification. Names from many Asian cultures, for instance, may not follow Western naming conventions, and it can be difficult to determine gender based solely on these names \cite{van2020gender,tien2023gendec}. 

\subsection{Demographic bias investigation research}
Investigating gender bias in scholarly research has far-reaching implications, as the outcomes of such studies can inform academic practices and guide the development or adjustment of policies aimed at promoting equality \cite{belingheri2021twenty, guthridge_promoting_2022}. The reliability of these studies fundamentally depends on the accuracy of the author name disambiguation and gender prediction methods or algorithms on which they are based. Various approaches to author name disambiguation and gender identification exist, including algorithmic solutions, name-based inference, manual verification, and the use of authors' self-reported data. 
 
Algorithmic solutions for gender identification, such as \href{https://genderize.io}{Genderize.io}, \href{https://gender-api.com}{Gender API}, \href{https://cran.r-project.org/web/packages/gender/index.html}{the gender package in R}, \href{https://github.com/ferhatelmas/SexMachine}{SexMachine in Python}, and \href{https://namsor.app}{Namsor}, rely on large datasets of name-to-gender associations. Tools like Genderize.io and Gender API predict gender by analyzing the frequency of names in databases and utilizing machine learning techniques. We refer the reader to the survey of these techniques by \cite{Santamaria2018NameGenderCompare} for details.

Name-based inference methods use predefined name lists or databases to infer gender, but their reliability can be limited due to cultural variability in naming conventions. Manual verification involves researchers assigning gender based on context, such as reviewing academic profiles, but this approach is time-intensive and prone to subjective biases. Alternatively, self-reported data, collected through surveys or directly from authors, offer the highest accuracy but may not always be available or complete. However, these methods exhibit significant variability in reliability and transparency, with many of them providing limited or inconsistent levels of accuracy \cite{santamaria2018comparison,lockhart2023name}.

\subsection{Causal factors influencing gender disparities}
Understanding the root causes of gender disparities in academia requires more than identifying gaps — it requires researchers to analyze the underlying variables that may contribute to these disparities \cite{mishra_self-citation_2018,hannak2020explaining}. To thoroughly examine gender disparities, it is critical to incorporate causal factors into the analysis to assess whether these factors account for or mitigate observed gender gaps. For example, factors such as career length (e.g., seniority), institutional rank, publication venue, and journal impact factor have been used to explore how structural or individual variables contribute to disparities \cite{liu_gender_2023,huang_historical_2020}. By including these variables, researchers can determine whether the disparities are reduced or explained away when considering specific contributing factors, providing a more nuanced understanding of gender bias. We observed that while studies we reviewed tended not to include potential causal factors in their analysis, recent papers increasingly incorporate these variables to add rigor to their conclusions. 

Some causal factors frequently included are authors' career lengths (i.e., seniority), year of publication, and number of authors. Furthermore, the papers studying causal factors tend to consider numerous variables, such as country, affiliation rank (e.g., whether the author is associated with a high-ranking or lower-ranking institution based on metrics such as global university rankings), venue of publication, institutional affiliation, academic roles (e.g., full, associate, or assistant professors), journal impact factors, and area of research. For instance, \citet{caplar_quantitative_2017} included seniority of the first author, number of references, total number of authors, year of publication, journal of publication, field of study, and geographical region of the first author's institution as causal factors into their analysis to observe the effect size of the gender bias after controlling the aforementioned causal factors. They found gender bias in their analysis, and controlling for causal factors makes their result claims more credible. Similarly, \citet{dworkin_extent_2020} and \citet{wang_gendered_2021} incorporated the year of publication, the number of authors, whether the paper was a review article, and the seniority of the paper's first and last authors into their analysis.

Papers that included causal factors show mixed results on the effect size, some discovering a decreased gender gap and some reporting no change in gender gap after controlling for causal factors. For instance, \citet{huang_historical_2020} reported that after controlling for career length as a causal factor, the gender gap in total productivity reduced from 31.0\% to 7.8\%, while including the country and affiliation rank did not further significantly affect the gender bias result, and considering total productivity as a causal factor eliminated the gender gap. \citet{dion2018gendered} discovered that more gender-diverse subfields and disciplines produce smaller gender citation gaps. However, \citet{odic_publication_2020} observed that male first authors achieve higher publication counts and citation counts even when affiliations are controlled for. Furthermore, they found no evidence for the assumption that the publication rate in individual subfields can account for the observed publication gender gap. This decrease in the gender gap after controlling for some causal factors implies that it is crucial to include causal factors in the analysis to add more rigor to the results. 

\subsection{Motivation for our work}
The choice and application of author name disambiguation and gender identification methods can significantly influence the results and their interpretation \cite{kim2016distortive}. For instance, \cite{kim2016distortive} 
showed how the academic community's reliance on the seemingly benign practice of using initials for author name disambiguation 
leads to significant distortions in the statistical properties of coauthorship networks, including underestimating the number of unique authors and network components and overestimating average productivity, number of co-authors per author, areas of expertise per author, and network density. Such errors caused by methodological flaws in author name disambiguation and gender identification underscore a) the profound influence that these inaccuracies can have on results and subsequent analyses, and b) the need for an in-depth assessment of methodological approaches and their implications on downstream research tasks. To address these needs, our study provides a comprehensive literature review that examines the range of techniques used for author name disambiguation and gender identification. This review spans a variety of disciplines, time periods, and locations, aiming to ensure robust, consistent, and objective findings.

By rigorously investigating these methodologies, our paper seeks to achieve three primary objectives:

\begin{itemize}
    \item A comprehensive analysis of the methods used for author name disambiguation and gender identification in 70 papers.
    \item The identification of common challenges and patterns in implementing author name disambiguation and gender identification methods.
    \item A standardization of the scholarly data analysis process through SoDA Cards as a solution to the outlined issues with variation and transparency in methods. 
\end{itemize}

\subsection{Scholarly Data Analysis (SoDA) Cards}
Standardized methods for benchmarking and documenting data provenance and measurement choices are integral to many fields \cite{gebru2021datasheets}. As a prime example, the machine learning community employs practices such as providing Model Cards \cite{mitchell2019model} and Datasheets \cite{gebru2021datasheets}. Model Cards provide detailed performance reports for trained machine learning models across various conditions, ensuring clear and human-readable communication about a model's capabilities, intent of use, and limitations. Similarly, Datasheets serve as a standardized means of conveying important characteristics of datasets used in machine learning. Although the machine learning community widely embraces transparent reporting practices, these methods can be overlooked in bibliometrics. 

Building upon the conventions of standardized practices in machine learning like Model Cards and Datasheets, this study proposes SoDA Cards, tailored for bibliometric research to foster transparency in reporting practices. While such reporting practices might appear to be predominantly suited for machine learning applications, they are equally relevant for bibliometric and scientometric studies because these fields heavily depend on author name disambiguation and gender identification methods and algorithms. The inherent complexity and variety of these methods accentuate the need for transparent and standardized reporting, which would foster greater reproducibility and comparability across studies. More importantly, it would pave the way for a more comprehensive understanding of the challenges and progress in the ongoing quest to understand and address gender and other demographic biases in the measurement and practices of the evolution of science and the scientific workforce. 

From our literature review, we learned that there are inconsistencies within and across methodological pipelines. In this section, we present our framework in the form of concise and standardized summaries documented on the "Scholarly  Data Analysis Card", which we designed as a potential solution to mitigate inconsistencies in adopting author name disambiguation and gender identification methods. These cards aim to standardize the reporting of methodologies to allow a variety of stakeholders of academia, e.g., universities, funders, policymakers, students, and algorithm developers, to compare results across individual studies and methodological practices that these studies used to arrive at their claims. 

\section{Patterns in scholarly data analysis for demographic bias}
The presented literature review followed a rigorous and systematic approach to ensure comprehensive coverage of relevant research. We utilized Google Scholar as our primary search engine to identify relevant literature. We used a set of keywords selected to maximize the breadth of the search, which we provide in Appendix \ref{sec:keywords}. Our sample of surveyed papers consisted of published articles that had undergone a peer-review process. We excluded unpublished and preprint papers (e.g., arXiv papers). Details of our paper selection are provided in the following sections.

\subsection{Paper search strategy and sampling criteria}
\label{sec:paper-sampling}

\begin{figure*}[!htbp]
    \centering
    \includegraphics[width=\textwidth]{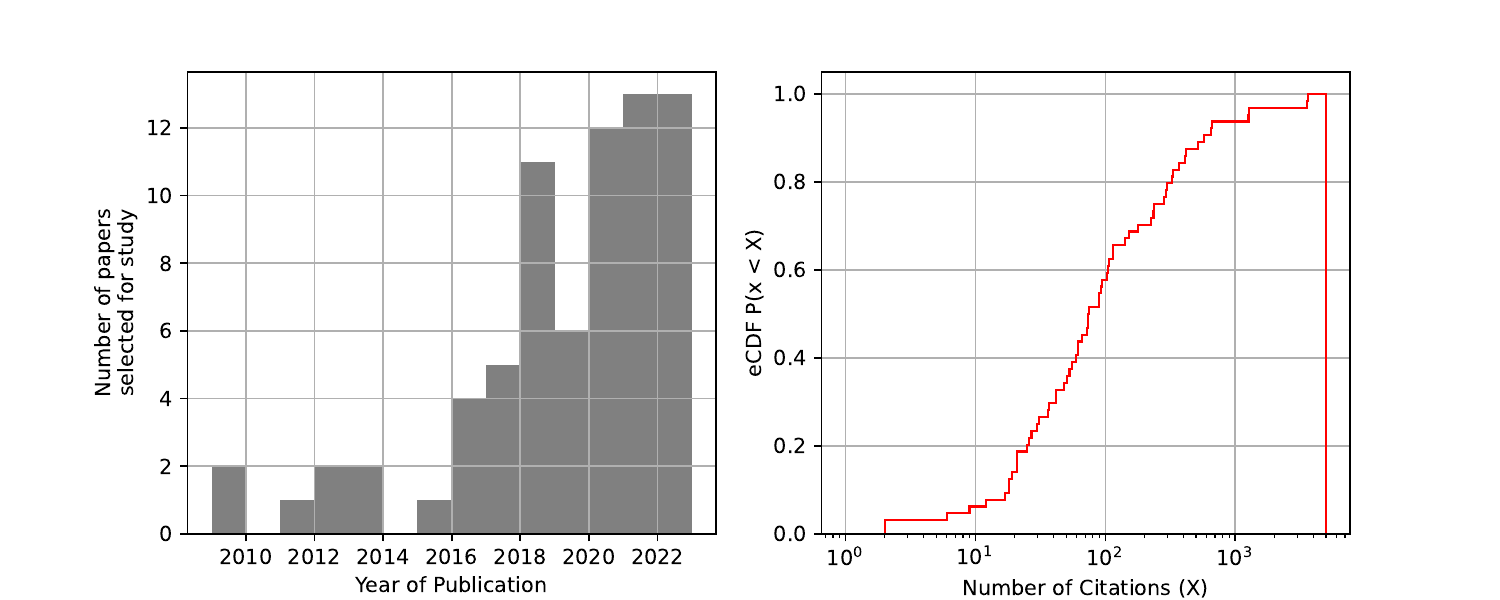}
    \caption{Distribution of selected papers}
    \label{fig:review-selection}
\end{figure*}

There has been significant research on various dimensions of demographic biases in scholarly work in recent years. We employed the following sampling method to retrieve papers likely to have influenced the study of demographic biases in scholarly research: 

\begin{itemize} 
    \item Identify papers to include in the review by searching Google Scholar with the keywords (in Appendix \ref{sec:keywords}). These keywords cover a wide selection of papers, e.g., on Google Scholar, ``gender bias scholarly analysis" led to 17K papers \footnote{\url{https://scholar.google.com/scholar?as_ylo=2009&as_yhi=2024&q=gender+bias+scholarly+data&btnG=}}. 
    \item We selected papers with high citation counts per keyword as reported by Google Scholar. This choice is motivated by the idea that highly cited papers can impact the methodology used in future papers in a field such that prioritizing highly cited papers allows us to detect dominant methodological practices in the field. We are aware that highly cited papers can be cited for reasons other than their methodology. However, our observation of these citations reveals that most citations of these influential papers utilize similar methodological practices as cited papers. This validates our assumption of using these papers to understand dominant practices in scholarly data analysis research.
    
    \item To ensure diversity in terms of time and venues of papers in our sample, we selected papers published between 2009 to 2023 from peer-reviewed journals and conferences.  
\end{itemize}

A chart illustrating the distribution of publication years and their corresponding citation counts is shown in Figure \ref{fig:review-selection}. It shows that most papers in our sample are from the recent past, primarily between 2018 and 2022. Our sample has a good diversity of papers in terms of citation counts as we have 30\% (\~20) papers with less than 30 citations.

\subsection{Annotation}
We annotated the papers in our sample, focusing on two key aspects: methods (developed and) used for author name disambiguation (AND) and gender identification. We first discuss how we annotated for AND, followed by gender identification. We identified five commonly used approaches for AND: no disambiguation, algorithmic methods, name-based approaches, manual searches, and self-reported data (gold standards). Each paper was categorized into one of these approaches, with detailed descriptions provided in a codebook (Table~\ref{tab:codebook}). 

To annotate gender identification methods, we developed a detailed codebook (see Table~\ref{tab:codebook}) to account for the diverse approaches used across various studies. Unlike author name disambiguation, which typically relies on a single method, gender identification often involves multiple strategies tailored to effectively identify names from different countries or cultural backgrounds. Consequently, our annotation process for gender identification differs from that of AND. Specifically, we evaluated each paper based on two distinct facets: the number of gender identification methods employed and the types of methods used. Initially, we identified the number of gender identification methods utilized in each paper, categorizing them as either “Single” or “Multiple.” This classification allowed us to capture whether a study relied on one method or incorporated several techniques for gender identification. Subsequently, we categorized each gender identification method into one of four types: algorithmic, self-reported, manual search, and name-based/heuristics. For example, if a paper exclusively employed algorithmic methods for gender identification, we annotated it as using a “Single” method and classified the method as “Algorithmic.” If a paper utilized multiple gender identification methods, we labeled it with each applicable method type, such as both “Algorithmic” and “Manual Search.” 

This annotation approach allows us to comprehensively understand the methodological landscape of gender identification in scholarly literature. By distinguishing between the number and types of gender identification methods, we can better analyze trends and preferences in research approaches, highlighting whether studies employ single-method approaches or integrate multiple techniques, e.g., to enhance accuracy and reliability. Detailed descriptions of each gender identification approach and the criteria for categorization are provided in Table~\ref{tab:codebook}. During the gender identification process, gender can be classified as a binary value or more than two categories. Some studies also incorporate "unknown" gender in the analysis. 

\begin{table*}[htbp]
\centering
\begin{tabular}{c|p{10cm}}  
\hline\hline
\textbf{Category} & \textbf{Definition}  \\ 
\hline
\multicolumn{2}{c}{\textbf{1. Author Name Disambiguation: Methods}}   \\  
\hline

No Disambiguation & Studies that 1) conduct analyses at the authorship level, 2) do not explain if and how they used any author name disambiguation methods, and 3) use data from Web of Science or Microsoft Academic Graph (MAG) without specifying whether the unique author identifiers provided in these datasets were used to distinguish between authors with similar or identical names. Simply relying on these datasets does not ensure disambiguation unless the identifiers are explicitly utilized to resolve ambiguities before analysis. \\  
Algorithm & Studies that use an algorithmic approach that utilizes name strings and additional features, such as affiliation, to disambiguate authors' names. \\  
Name-based (heuristics) & Studies that disambiguate author names based on name strings (e.g., first name initials or all initials)   \\ 
Manual search & Studies in which human annotators search the web for images and institutional web pages to disambiguate the authors' names.\\  
Gold standards & Studies that use survey data, specifically self-reported data. 
\\  
\hline
\multicolumn{2}{c}{\textbf{2. Gender Identification: Number of Methods Used}}   \\  
\hline
Single Method & Studies that used only one method to identify the gender identities of the authors.  \\  
Multiple Methods & Studies that used more than one method to identify the gender identities of the authors.  \\  \hline
\multicolumn{2}{c}{\textbf{3. Gender Identification: Methods}}   \\  \hline
Gold standards & Studies that rely on self-reported gender identities provided by the authors. \\  
Manual search & Studies that rely on external sources (e.g., SSN or database), photographs, titles (e.g., Mr., Mrs.), or pronouns (e.g., he, him) to identify the gender identities of the authors. \\  
Heuristics & Studies that use only authors' name strings (e.g., Genderize.io, gender guesser), to identify the gender identities of the authors. \\  
Algorithm & Studies that use algorithms that use more than just the authors' name strings to identify the gender identities of the authors, this includes year of birth, location, e.g. Genni 2.0+Ethnea \cite{torvik2016ethnea}.  \\  \hline\hline
\end{tabular}
\caption{Summary of codebook definitions for categorizing papers (here referred to as studies)}
\label{tab:codebook}
\end{table*}

\subsection{Evaluation of annotations using kappa scores}
We calculate Cohen's kappa ($\kappa$) to measure the interrater reliability of annotating author name disambiguation and gender identification methods \cite{cohen1960coefficient}. Cohen's kappa can range from -1 to 1, where 0 represents the amount of agreement that can be expected by chance, and 1 represents perfect agreement between the raters. If $Pr(a)$ gives the proportion of times two raters agree, and $Pr(e)$ gives the proportion of expected agreement, then $\kappa$ can be represented by:

\begin{equation}
\label{eq:kappa}
\text{Cohen's kappa} (\kappa) = \frac{Pr(a) - Pr(e)}{1 - Pr(e)}
\end{equation}

This study employed three doctoral researchers as annotators in a collaborative data coding process. A corpus of 70 papers were analyzed with respect to three key aspects: the Author Name Disambiguation methods implemented, the number and specific type(s) of Gender Identification methods utilized, and instances of gender identification performed using a single method. To ensure methodological transparency, inter-annotator agreement was assessed prior to the resolution of discrepancies, with corresponding Cohen’s kappa coefficients reported for each aspect, as detailed below.

\begin{itemize} 
    \item Author Name Disambiguation: 0.81
    \item Gender Identification (Number of methods): 0.85
    \item Gender Identification (Single): 0.71
\end{itemize} 

We consider scores in the range of 0.71 – 0.86 to indicate moderately strong agreement \cite{sim2005kappa,mchugh2012interrater}, reflecting a generally consistent annotation process. However, disagreements between annotations were resolved through discussion, and the original annotations were adjusted to reflect the consensus, ultimately achieving 100\% agreement.

\section{Results}
The results are organized into subsections that discuss key themes and insights from the review process, focusing on author name disambiguation and gender identification. 

\subsection{Author Name Disambiguation}
This subsection outlines the distribution of the disambiguation methods used in the papers that are included in our sample. Additionally, we discuss how these studies addressed the challenge of disambiguating Asian names, highlighting their approaches and implications.

\subsubsection{Analysis of author name disambiguation methods}
In this section, we first present the distribution of the AND methods used in the 70 papers in our sample (\ref{tab:GI_papers}), and then explain potential issues with each method. We found that 51.4\% (36/70) of the analyzed papers did not use any author name disambiguation (i.e., authorship-level analysis). Among those that employed disambiguation methods, 21.4\% (15/70) used algorithmic disambiguation, 12.9\% (9/70) name-based (heuristic) approaches, 10.0\% (7/70) manual search methods, and 4.3\% (3/70) gold-standard author identity data. The precision of AND is essential for conducting analyses at the individual level, rather than just aggregating authorship-level data. However, current practices fall short of achieving this precision. One potential issue is the prevalent underutilization of AND methods, leading to substantial limitations of research analysis results and conclusions. The failure to accurately identify individual authors can narrow the scope of analysis as it may exclude critical causal factors. For a comprehensive and nuanced understanding, it is crucial to account for factors such as an author's career duration, publication history, and field of work \cite{mishra_self-citation_2018}. This level of detail requires a thorough disambiguation of each author's identity, moving beyond authorship-level analysis. 

The practice of relying on methods for AND, such as using full names or combinations of initials with last names, often proves inadequate and leads to errors in correctly identifying authors: taking a name-based approach becomes problematic in cases of name overlap or inconsistent indexing. The resulting errors are the merging of distinct authors with similar names into a single identity and the splitting of a single author's papers into multiple identities \cite{kim2016distortive}. Such inaccuracies can distort collaboration networks and misattribute scholarly output, which may undermine the reliability of research findings, particularly in gender bias studies that seek to accurately track the contributions and impact of individual researchers. Furthermore, the lack of widespread implementation of robust methods, such as algorithmic disambiguation or the use of gold-standard data, is a limitation in many studies that utilize author name disambiguation. The prevailing reliance on less accurate, heuristic approaches highlights a pressing need for the adoption of more sophisticated and reliable methods in author name disambiguation. This is particularly critical in research analytics and bibliometrics, where the precise attribution of work is essential.

\begin{figure}[htp]
\centering
\includegraphics[width=8cm]{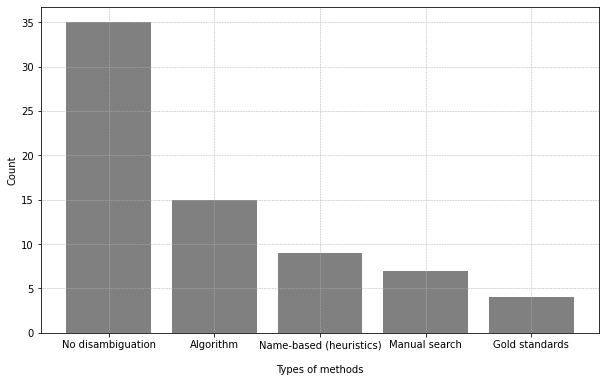}
\caption{Author Name Disambiguation Methods Distribution Barchart}
\label{fig:and_dist}
\end{figure}

\subsubsection{Overview of elaboration on Asian names}
Some studies have examined how names are treated across different cultural and geographic regions by specifying the countries included in their analyses. For example, \citet{huang_historical_2020} and \citet{mishra_self-citation_2018} listed the countries covered in their research. \citet{huang_historical_2020}, in particular, considered 83 countries in their country-specific analysis. This level of detail is crucial for understanding the geographic and cultural scope of their analyses, especially when dealing with the diversity and complexity of naming conventions. Names can vary significantly across regions in terms of structure, order of family and given names, and the frequency of certain surnames \cite{mcelduff2008s}. Such differences require researchers to adopt culturally sensitive methodologies to ensure accurate identification and categorization. By providing transparent documentation of the countries included in their studies, researchers enhance the interpretability of their findings and allow readers to assess the extent to which their methods account for cultural variations.

\begin{table*}[htbp]
\centering
\begin{tabular}{c|c|p{8cm}}  
\hline\hline
\textbf{Category} & \textbf{Percentage} & \textbf{List of papers} \\ \hline
No Disambiguation & 36 out of 70 ($51.43\%$) &  
\cite{chatterjee_gender_2021,lariviere_bibliometrics_2013,king_pandemic_2021,huang_historical_2020,maliniak_gender_2013, yalamanchali2021trends, jemielniak2023covid, wu2020gender, palser2022gender, mohammad-2020-gender, dion2018gendered, holman2018gender,gonzalez2019contemporary,thomas2019gender,vranas2020gender,edwards2018gender, squazzoni2021peer, bendels2018gender,card2020referees, filardo2016trends, gayet2019comparison,lerchenmueller2019gender,west2013role,ostby2013gender, yang2022gender,copenheaver2010lack,benjamens2020gender, schisterman2017changing,squazzoni2021gender,qamar2020gender,fox2018patterns,hagan2020women, nunkoo2020three, salerno2019male, merriman2021gender,thomas2019gender}\\  
Algorithm & 15 out of 70 ($21.43\%$) &  
\cite{pilkina_gender_2022, liu_gender_2023,ross_women_2022,pinheiro_women_2022,abramo_gender_2009,lariviere_sex_2011,nielsen_gender_2016,van2017vicious,zhao_gender_2023,mishra_self-citation_2018,mihaljevic2016effect,kozlowski2022intersectional,zhang2022gender,fulvio2021gender,murphy2020open}  \\  
Name-based (heuristics) & 9 out of 70 ($12.86\%$) &  
\cite{caplar_quantitative_2017, dworkin_extent_2020, wang_gendered_2021,teich_citation_2022,king_men_2017,odic_publication_2020,kong2022influence,vasarhelyi_gender_2021,maggio2023voices}  \\  
Manual search & 7 out of 70 ($10.00\%$) &  
\cite{azoulay_self-citation_2020,mayer2018does,mcdermott2018sex,carter2017gender,van2016gender,ni2021gendered,ahmadia2021limited}  \\  
Gold standards & 3 out of 70 ($4.28\%$) &  
\cite{eloy2013gender, tao2017gender,raj2016longitudinal}  \\    \hline \hline
\end{tabular}
\caption{Author Name Disambiguation Methods Paper Categorization}
\label{tab:author}
\end{table*}

\subsection{Gender Identification}
\subsubsection{Overview of gender identification methods}
In this section, we present the result of the distribution of gender identification methods in the papers in our corpus. Compared to the author name disambiguation methods, which usually involve a single method for disambiguating authors' names, numerous studies relied on multiple methods for identifying authors' gender identities. We found that 45/70 ($64.3\%$) \cite{pilkina_gender_2022,caplar_quantitative_2017,lariviere_bibliometrics_2013,liu_gender_2023,ross_women_2022,teich_citation_2022,pinheiro_women_2022,king_pandemic_2021,huang_historical_2020,abramo_gender_2009,lariviere_sex_2011,yalamanchali2021trends,jemielniak2023covid,wu2020gender,palser2022gender,king_men_2017,mishra_self-citation_2018,odic_publication_2020,mayer2018does,dion2018gendered,gonzalez2019contemporary,mcdermott2018sex,bendels2018gender,card2020referees,carter2017gender,gayet2019comparison,lerchenmueller2019gender,eloy2013gender,west2013role,van2016gender,tao2017gender,vasarhelyi_gender_2021,copenheaver2010lack,benjamens2020gender,kozlowski2022intersectional,carr2018gender,raj2016longitudinal,ni2021gendered,qamar2020gender,ahmadia2021limited,hagan2020women,maggio2023voices,murphy2020open,nunkoo2020three,salerno2019male} of the papers used a single gender identification method, whereas 25/70 ($35.7\%$) of the papers \cite{chatterjee_gender_2021,dworkin_extent_2020,wang_gendered_2021,maliniak_gender_2013,azoulay_self-citation_2020,nielsen_gender_2016,van2017vicious,zhao_gender_2023,mohammad-2020-gender,holman2018gender,edwards2018gender,squazzoni2021peer,mihaljevic2016effect,card2020referees,filardo2016trends,kong2022influence,yang2022gender,schisterman2017changing,squazzoni2021gender,zhang2022gender,thomas2019gender,vranas2020gender,fulvio2021gender,fox2018patterns,merriman2021gender} used more than one method for identifying authors' gender identities. Moreover, even within the studies using multiple approaches, we observed that the order in which they used each method and the combination of multiple methods varied. To accurately display the distribution of gender identification methods used across the 70 reviewed papers, we counted the frequency of each method (Table \ref{tab:GI_papers}). This count includes the methods (e.g., algorithm, heuristics, manual search, and gold standard) used in papers we reviewed, with special consideration for papers employing multiple approaches. Each method was counted individually in these cases, allowing a single paper to contribute to multiple method categories. This approach ensures a detailed and nuanced representation of the methods' distribution, acknowledging the complexities inherent in multi-method research papers.

\begin{table*}[htbp]
\centering
\begin{tabular}{p{2.35cm}|c|p{4.35cm}|p{4.35cm}}  
\hline\hline
\textbf{Category} & \textbf{Percentage} & \centering\textbf{Single method} & \textbf{Multiple methods} \\ \hline

Algorithm & 18 out of 98 ($18.37\%$) &  
\cite{liu_gender_2023,ross_women_2022,teich_citation_2022,pinheiro_women_2022,king_pandemic_2021,yalamanchali2021trends,mishra_self-citation_2018,bendels2018gender,murphy2020open}  
&\cite{dworkin_extent_2020,wang_gendered_2021,zhao_gender_2023,squazzoni2021peer,kong2022influence,yang2022gender,squazzoni2021gender,thomas2019gender,merriman2021gender}\\  

Heuristics & 41 out of 98 ($41.84\%$) &  
\cite{pilkina_gender_2022,caplar_quantitative_2017,lariviere_bibliometrics_2013,huang_historical_2020,jemielniak2023covid, king_men_2017,odic_publication_2020, dion2018gendered,
gonzalez2019contemporary,carter2017gender,gayet2019comparison,lerchenmueller2019gender,west2013role,vasarhelyi_gender_2021,benjamens2020gender,kozlowski2022intersectional,hagan2020women,maggio2023voices} 
&\cite{chatterjee_gender_2021,dworkin_extent_2020,wang_gendered_2021,maliniak_gender_2013,azoulay_self-citation_2020,van2017vicious,zhao_gender_2023,mohammad-2020-gender,holman2018gender,edwards2018gender,squazzoni2021peer,mihaljevic2016effect,card2020referees,filardo2016trends,kong2022influence,yang2022gender,schisterman2017changing,squazzoni2021gender,zhang2022gender,thomas2019gender,vranas2020gender,fulvio2021gender,fox2018patterns}\\

Manual search & 22 out of 98 ($22.45\%$) &  
\cite{mayer2018does,mcdermott2018sex,qamar2020gender,ahmadia2021limited,nunkoo2020three,salerno2019male}  
& \cite{chatterjee_gender_2021,maliniak_gender_2013,azoulay_self-citation_2020,nielsen_gender_2016,van2017vicious,holman2018gender,edwards2018gender,mihaljevic2016effect,card2020referees,filardo2016trends,schisterman2017changing,zhang2022gender,vranas2020gender,fulvio2021gender,merriman2021gender,fox2018patterns}\\  

Gold standards & 17 out of 98 ($15.46\%$) &  
\cite{abramo_gender_2009,lariviere_sex_2011,wu2020gender,palser2022gender,eloy2013gender,ostby2013gender,van2016gender,tao2017gender,copenheaver2010lack,carr2018gender,raj2016longitudinal,ni2021gendered}   
& \cite{maliniak_gender_2013,nielsen_gender_2016,mohammad-2020-gender,
azoulay_self-citation_2020,schisterman2017changing}\\  

\hline\hline
\end{tabular}
\caption{Categorizing Papers on Gender Identification Methods}
\label{tab:GI_papers}
\end{table*}

We found that name-based (heuristics) was the most frequently used approach among all gender identification methods ($42.27\%$ of papers). This approach involves using authors' first names to identify associated gender identities. For instance, Genderize.io is also included in this gender identification category, as it relies on authors' name strings to determine gender identities, though the specific method it employs may vary. The name-based approach has limitations as many names are unisex, and map to different genders depending on the culture or the country of affiliation (e.g., Andrea is typically a male name in Italy and related cultures but is a female name in the U.S. and English cultures, both NamSor and Genderize.io assign it the female gender), and are dependent on the full name (e.g., Harpreet is a unisex name in the Punjab region of India, but the gender can often only be identified after knowing the full name, e.g., Harpreet Kaur is female while Harpreet Singh is more likely to be male). On the other hand, the Gold standard approach, notable for its ideal approach of relying on self-reported gender identity, was utilized least frequently, accounting for only ($15.46\%$) of papers. Other methods, namely manual search ($23.71\%$) and algorithm-based ($18.56\%$), were also used.

\begin{figure}[htp]
\centering
\includegraphics[width=8cm]{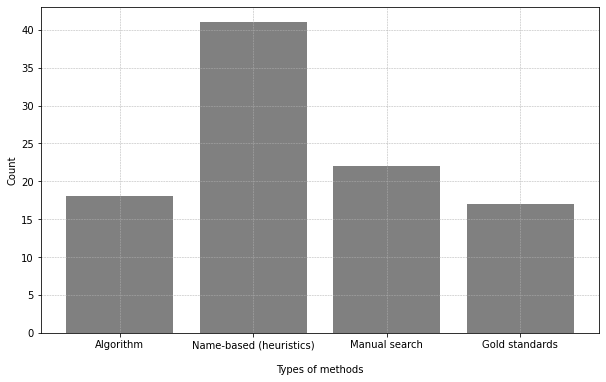}
\caption{Distribution of Gender Identification Methods}
\label{fig:gi_dist}
\end{figure}

\subsubsection{Use of gender labels and handling of unknown gender identities}
Gender categories in the identification process are typically binary (i.e., male and female), which were also the most commonly observed identities in our sample. However, the reliance on binary labels for gender categorization presents a significant limitation as it restricts the acknowledgment of diverse gender identities, including non-binary, by forcing them into one of two predefined categories \cite{you-etal-2024-beyond}. Further, this binary constraint hinders a more comprehensive understanding of gender bias at a fine-grained level. The complexity of this issue is compounded by the fact that gender identification algorithms predominantly use binary labels, thus perpetuating this narrow classification framework. Nevertheless, some studies have attempted to move beyond binary labels to provide a more comprehensive view of gender inequalities in academia. For instance, \citet{lariviere_bibliometrics_2013} used binary categories and unisex labels for the names appearing in both lists as unisex. 

Additionally, we noticed that some studies introduced a separate `Unknown' category for authors whose gender identities could not be determined \cite{pinheiro_women_2022,mishra_self-citation_2018,mohammad-2020-gender,squazzoni2021peer,vasarhelyi_gender_2021}. This approach, particularly the effort to acknowledge rather than disregard authors with unassigned gender by including a specific label for gender-unidentified authors, represents a step towards inclusivity and a more nuanced understanding of gender dynamics in academia. We also noted a methodological concern: numerous studies tend to exclude authors with indeterminate gender identities when analyzing gender disparities. This approach may have drawbacks, particularly in the context of Asian names, such as those of Chinese or Korean origin, where determining gender identity is often more challenging. For instance, excluding authors whose gender cannot be ascertained disproportionately impacts different ethnicities differently, potentially skewing the sample representation and further biasing the study outcomes. This exclusionary practice also diminishes the inclusivity of research, as it overlooks valuable contributions from gender-unclassified authors, thereby underrepresenting the diversity of scholarly efforts. Therefore, this method risks perpetuating biases, ethnic bias, as certain groups are more likely to be excluded due to cultural naming conventions; and systemic bias, where existing methodologies fail to evolve with changing societal and academic norms. These biases collectively limit the scope of analysis, hinder equitable representation, and potentially obscure critical insights into the intersections of identity and academic contributions. 

\subsubsection{Considerations of Asian names}
Our literature review highlighted challenges in handling Asian-origin names, particularly in the context of gender identification. The prevalence of unisex names for scholars from these regions makes it difficult to accurately determine the gender of authors with Asian names. We observed that some studies opted to exclude authors with Asian names from their analyses. This exclusion, which impacts authors from certain ethnic groups, is sometimes explicitly mentioned in the papers and introduces biases in the analysis. For instance, overlooking influential scholars with high citation counts from specific fields can distort the representation of academic impact and productivity. In studies focusing on citation counts or publication volumes, such exclusions may lead to incomplete or misleading conclusions about disparities in academic contributions across genders and ethnic groups. 

Moreover, we noticed that most papers lacked detailed information about how names from different regions, particularly Asian countries, were handled. Specifically, there's often little to no information on the proportion of names from Asian countries or the differential treatment of authors with Asian origins. However, some studies have addressed this issue directly. For instance, \citet{huang_historical_2020} provided a comprehensive explanation for their decision to exclude certain ethnic groups from their analysis: they acknowledged that while many name disambiguation algorithms effectively help to reconstruct the careers of authors with European names, they struggle with those of Asian origin. This, combined with the difficulty in inferring gender from Asian names, led them to exclude researchers from China (including mainland, Hong Kong, Macau, and Taiwan), the Democratic People's Republic of Korea (North Korea), Japan, Malaysia, the Republic of Korea (South Korea), and Singapore \cite{huang_historical_2020}. Such disclosures are vital as they inform readers about potential limitations to the generalizability of a study's results.

\section{Scholarly Data Analysis (SoDA) Cards}
In this section, we explain how we developed the concept of SoDA Cards and how we suggest that researchers and practitioners use them. We acknowledge that relying on the authors' self-reported data, which is the most accurate method, becomes impractical when working with large datasets. Our review of the literature has revealed that many studies adopt alternative methods for author name disambiguation and gender identity determination. Especially, we observed the lack of clear descriptions of implementation details for both tasks, which complicates comparisons between studies. To address this issue in future research, we introduce the SoDA Cards.

The SoDA Cards are designed to provide comprehensive and concise information about study objectives and the data analysis process. These cards entail several subsections, including study specifications, corpus profiling, and detailed descriptions of the methods used for author name disambiguation and gender identification. The cards also cover the causal factors and methodologies used to analyze gender bias, along with the results of these analyses. We provide an unfilled template of the Scholarly Analysis Card in Figure \ref{fig:unfilled_card} and a filled card for illustrative purposes in Figure \ref{fig:filled_card}. The "Scholarly Data Analysis Card" enables researchers to transparently and accessibly document their analytical workflows, enhancing reproducibility in the field of bibliometrics. The SoDA Cards are available in the GitHub repository: \url{https://github.com/HaeJinLee41/scholarly_bias_study}.

\subsection{Study specification}
In the study specification section, authors are encouraged to clearly define the performance metrics used in their study, such as the number of citations, author composition, impact factors, and so on. Additionally, they should indicate whether the study's code or data is publicly accessible and, if so, provide relevant links or references. 

\subsection{Corpus profile}
In this section, authors are expected to provide details about the dataset used in their study.

\textbf{Data source(s)}: The source(s) of the dataset, including whether they merged data from multiple sources. Furthermore, authors should document the data sources used with data citations. 

\textbf{Domain}: Whether the dataset represents a specific domain or field, or if it encompasses interdisciplinary perspectives to gain broader insights into gender bias across various disciplines. 

\textbf{Corpus volume}: The initial and final size of the dataset used for analysis.

\textbf{Geographic scope}: The geographical focus of their study. Is it exploring gender bias in a specific country or a particular continent, or does it aim to understand broader bias trends across multiple countries? 

\textbf{Temporal Scope}: The period of the dataset should be clearly stated. 

\subsection{Author name disambiguation}
Comprehensive information about the author name disambiguation methods they used and the methods used to evaluate it. 

\textbf{Disambiguation method}: The techniques used for author name disambiguation. This could include heuristic approaches, specific algorithms, or manual searches. If multiple methods were used, all should be detailed. In cases where disambiguation was not performed, an explanation is required. For example, reliance on self-reported data might justify the absence of disambiguation. 

\textbf{Evaluation method}: If the effectiveness of the disambiguation technique was assessed, authors should describe the methods employed for this evaluation.

\textbf{Evaluation data}: The dataset(s) used to assess the disambiguation method, whether a random sample check or a different dataset.

\textbf{Accuracy}: If the disambiguation method was evaluated, the achieved accuracy should be stated. 

\textbf{Handling on names across diverse cultural and ethnic backgrounds}:
Authors should clarify how they addressed variations in author names from diverse cultural and ethnic backgrounds. This includes considering naming conventions, such as the placement of family names and prevalent surnames, and whether specialized algorithms or datasets for specific cultural or linguistic contexts were used.

\subsection{Gender idenitfication}
Detailed information on the methods used for gender identification, including any limitations faced, should be outlined.

\textbf{Gender identification method}: 
The approaches used for determining gender identities and the number and type of methods (algorithms, gold standards, etc.) employed. 

\textbf{Name part used for gender identification}: Which part of the name (first, last) was utilized to identify gender. 

\textbf{Gender categories used}: The labels for gender categories. This includes any unique categories, like 'unidentified'.

\textbf{Percentage of unidentified gender}: 
The proportion of gender identities that remained unidentified.

\textbf{Handling of unknown gender}: 
How data from authors with unidentified gender identities was managed - whether they were excluded or categorized separately. 

\textbf{Evaluation method, data, and accuracy}:
The evaluation method, data, and accuracy of the gender identification process. 

\textbf{Handling of names across different ethnic groups for gender identification}:
for names, including specific considerations or methodologies used to account for variations across different ethnic or cultural groups. For instance, authors could elaborate on how they handled names of Asian origin (e.g., did they disregard them or use a different approach for this group?).

\subsection{Analysis}:
The methods used to analyze gender bias, including relevant specifics, causal factors considered, and the analytical pipeline. 

\textbf{Method}:
The methods they employed in their gender bias analysis. If multiple methods were used, each should be thoroughly described. 

\textbf{Control for causal factors}: 
Whether the analysis accounted for factors like the year of publication, the author's career age, the field of work, and prior publication history.

\subsection{Results}:
The findings of the study, with a particular emphasis on results related to gender bias. 

\textbf{Presence of gender bias}:
Whether their analysis revealed any gender bias. If the findings were insignificant, they should provide insights into potential reasons. 

\textbf{Overall effect size and effect size post-control}:
The overall effect size that was discovered in the study. Additionally, authors need to detail the effect size after considering various causal factors. This includes explaining how the results were influenced by adjustments for specific factors and if such adjustments were made in the analysis.

\section{Limitations}

This study presents a framework for accurate scholarly data analysis; however, both the findings derived from prior research and the proposed framework have limitations. Regarding the analysis of existing scholarly literature, while the results are considered representative, they are subject to limitations imposed by the sample size. The papers were selected based on representative keywords related to scholarly data analysis and demographic bias, identified through standard search terms. While this approach captures prominent research in the field, it may exclude relevant publications that do not explicitly feature these keywords or are not readily indexed by Google Scholar. Given the prevalence of the chosen keywords, this potential omission is not expected to impact the overall findings significantly. Furthermore, the selection process prioritized highly cited papers (as detailed in Section \ref{sec:paper-sampling}). These influential papers are frequently referenced for their methodological contributions, suggesting that subsequent studies adopting similar methodologies will likely exhibit comparable characteristics to those identified in our analysis.

The proposed framework also has inherent limitations. Specifically, the emphasis on demographic attributes such as gender may pose challenges in contexts where these attributes cannot be reliably inferred due to privacy concerns or local regulations. However, this limitation is mitigated by the understanding that datasets with such restrictions should not be used for aggregated demographic scholarly data analysis in the first place. Therefore, the framework's focus on transparency regarding gender assignment methodologies can contribute to compliance with relevant policies. While not a primary objective, this potential framework application warrants further investigation.

\section{Discussion and Conclusion}
Our study highlights the critical need to document and refine author name disambiguation and gender identification practices in scholarly research. Improving these processes is essential to enhance the accuracy and reliability of studies investigating gender biases. To conduct reliable investigations of gender biases at the authors' career level, it is vital to adhere to key methodological steps. These include rigorous author identification and disambiguation and the accurate attribution of demographic details such as gender, race, and nationality. Moreover, it's crucial to control for causal factors that might impact the variables under study. These essential steps are illustrated in Figure \ref{fig:flow_chart}. Inaccuracies in author identification not only compromise the integrity of data but also lead to skewed analyses, which can yield biased interpretations of gender disparities in academia. This hinders a comprehensive understanding of the factors driving these disparities and might adversely impact policymaking. 

Given these observations, our findings have significant implications for future research in this domain. Emphasizing precise author attribution and acknowledging individual contributions become pivotal in studies exploring gender bias. As the field progresses, developing more sophisticated and reliable methods for author name disambiguation is essential to validate findings in gender bias research. To move forward, we propose the introduction of a model card system for bibliometric studies. These SoDA Cards serve as a standardized tool for researchers to document and share their methodologies and results. They include detailed information about the data, analytical methods employed, and the controls implemented for causal factors. Such a system would not only promote transparency and replicability in gender bias research but also encourage the adoption of best practices across the field, ultimately leading to more accurate and insightful findings.



\providecommand{\tightlist}{%
  \setlength{\itemsep}{0pt}\setlength{\parskip}{0pt}}

\begin{figure}[htp]
\centering
\fbox
{
\begin{minipage}{0.9\linewidth}
\titlespacing{\section}{0pt}{\parskip}{-\parskip}
\titlespacing{\subsection}{0pt}{\parskip}{-\parskip}
\titlespacing{\subsubsection}{0pt}{\parskip}{-\parskip}
\titleformat{\section}{\small\fontsize{10}{12}\bfseries}{\thesection}{0.5em}{}
\setcounter{section}{0}
\small
\section{Paper Title}\label{paper-title}

Quantitative evaluation of gender bias in astronomical publications from
citation counts (Caplar et al., 2017)

\section{Study Specification}\label{study-specification}

\begin{itemize}
\tightlist
\item
  \textbf{Performance metric(s)}: number of citations
\item
  \textbf{Code/Data accessibility}: code and data are accessible
\end{itemize}

\section{Corpus Profile}\label{corpus-profile}

\begin{itemize}
\tightlist
\item
  \textbf{Domain}: astronomy
\item
  \textbf{Data source}: SAO, NASA, ADS, arXiv
\item
  \textbf{Corpus volume}: 149,741 papers
\item
  \textbf{Geographical scope}: global-level
\item
  \textbf{Temporal scope}: 1950-2015
\end{itemize}

\section{Author Name Disambiguation}\label{author-name-disambiguation}

\begin{itemize}
\tightlist
\item
  \textbf{Disambiguation method}: name-based method
\item
  \textbf{Evaluation method}: did not evaluate
\item
  \textbf{Evaluation data}: not applicable
\item
  \textbf{Accuracy}: not applicable
\item
  \textbf{Elaboration on names with Asian origins}: not applicable
\end{itemize}

\section{Gender Identification}\label{gender-identification}

\begin{itemize}
\tightlist
\item
  \textbf{Gender identification method}: algorithmic
\item
  \textbf{Name part used for gender identification}: first name
\item
  \textbf{Gender categories used}: binary (female and male)
\item
  \textbf{Percentage of unidentified gender}: 1.5\% (2260)
\item
  \textbf{Handling of unknown gender}: excluded from analysis
\item
  \textbf{Evaluation method}: did not evaluate
\item
  \textbf{Evaluation data}: not applicable
\item
  \textbf{Accuracy}: not applicable
\item
  \textbf{Elaboration on names with Asian origins}: not applicable
\end{itemize}

\section{Analysis}\label{analysis}

\begin{itemize}
\tightlist
\item
  \textbf{Method}: machine learning (random forest model)
\item
  \textbf{Control for year of publication?}: yes
\item
  \textbf{Control for author career age?}: yes (seniority)
\item
  \textbf{Control for field of work?}: yes
\item
  \textbf{Control for author prior publication?}: no
\end{itemize}

\section{Results}\label{results}

\begin{itemize}
\tightlist
\item
  \textbf{Presence of gender bias}: yes
\item
  \textbf{Overall effect size}: men received around 6\% more citations
  on average then women (considering year)
\item
  \textbf{Effect size after controlling causal factors}: papers authored
  by women receive 10.4 ± 0.9\% fewer citations than would be expected
  if the papers with the same non-gender-specific properties were
  written by men
\end{itemize}
\end{minipage}
}
\caption{Model Card example}
\label{fig:filled_card}
\end{figure}



\begin{figure}[htp]
\centering
\fbox
{
\begin{minipage}{0.9\linewidth}
\titlespacing{\section}{0pt}{\parskip}{-\parskip}
\titlespacing{\subsection}{0pt}{\parskip}{-\parskip}
\titlespacing{\subsubsection}{0pt}{\parskip}{-\parskip}
\titleformat{\section}{\small\fontsize{10}{12}\bfseries}{\thesection}{0.5em}{}
\setcounter{section}{0}
\small
\section{Paper Title}\label{paper-title}

\section{Study Specification}\label{study-specification}

\begin{itemize}
\tightlist
\item
  \textbf{Performance metric(s)}:
\item
  \textbf{Code/Data accessibility}:
\end{itemize}

\section{Corpus Profile}\label{corpus-profile}

\emph{Information about data}

\begin{itemize}
\tightlist
\item
  \textbf{Domain}:
\item
  \textbf{Data source}:
\item
  \textbf{Corpus volume}:
\item
  \textbf{Geographical scope}:
\item
  \textbf{Temporal scope}:
\end{itemize}

\section{Author Name Disambiguation}\label{author-name-disambiguation}

\emph{Information about author name disambiguation process, evaluation
method, and outputs}

\begin{itemize}
\tightlist
\item
  \textbf{Disambiguation method}:
\item
  \textbf{Evaluation method}:
\item
  \textbf{Evaluation data}:
\item
  \textbf{Accuracy}:
\item
  \textbf{Elaboration on names with Asian origins}:
\end{itemize}

\section{Gender Identification}\label{gender-identification}

\emph{Information about author gender identification process, evaluation
method, and outputs}

\begin{itemize}
\tightlist
\item
  \textbf{Gender identification method}:
\item
  \textbf{Name part used for gender identification}:
\item
  \textbf{Gender categories used}:
\item
  \textbf{Percentage of unidentified gender}:
\item
  \textbf{Handling of unknown gender}:
\item
  \textbf{Evaluation method}:
\item
  \textbf{Evaluation data}:
\item
  \textbf{Accuracy}:
\item
  \textbf{Elaboration on names with Asian origins}:
\end{itemize}

\section{Analysis}\label{analysis}

\emph{Information about causal factors and analysis}

\begin{itemize}
\tightlist
\item
  \textbf{Method}:
\item
  \textbf{Control for year of publication?}:
\item
  \textbf{Control for author career age?}:
\item
  \textbf{Control for field of work?}:
\item
  \textbf{Control for author prior publication?}:
\end{itemize}

\section{Results}\label{results}

\emph{Information about study results}

\begin{itemize}
\tightlist
\item
  \textbf{Presence of gender bias}:
\item
  \textbf{Overall effect size}:
\item
  \textbf{Effect size after controlling causal factors}:
\end{itemize}

\end{minipage}
}
\caption{Model Card template}
\label{fig:unfilled_card}
\end{figure}

\pagebreak

\bibliography{citations}

\appendix
\section{Appendix}
\subsection{Paper sampling criteria} \label{sec:keywords}
In this subsection, we list the keywords used to sample papers for this study. To capture a wide range of discussions on gender biases and related scholarly performance, we combined terms such as “gender bias,” “gender gap,” “gender disparities,” “gender differences,” “publication/academic/scientific performance,” “citation(s)/gendered citation,” “demographic bias,” “productivity,” and “scholarly analysis.” These keywords were applied individually and in combination to maximize the breadth and relevance of our literature search, while minimizing redundancy. For example, we paired “gender gap” with “citations” or “gender bias” with “productivity” to capture a variety of relevant contexts. 

\subsection{Overview of trends in gender bias research}
We analyzed themes and trends across domains, data sources, geographical and temporal data scopes, and scholarly performance indicators for the papers in our sample. We note that we do not provide specific percentages or proportions of papers for these dimensions, as our emphasis is on uncovering general trends rather than presenting granular quantitative breakdowns.  

\subsubsection{Domain} 
Gender biases refer to the systemic favoring of one gender over others, often resulting in unequal opportunities, representation, or treatment in societal, academic, or professional contexts. Gender bias has been studied in a wide range of fields. For instance, numerous studies in medicine \cite{chatterjee_gender_2021, filardo2016trends, gayet2019comparison} used bibliometric analyses to detect gender-based disparities in publication patterns. Similar large-scale data analyses have been conducted in astronomy 
\cite{caplar_quantitative_2017, wang_gendered_2021,dworkin_extent_2020,mcdermott2018sex}, physics \cite{teich_citation_2022} and psychology \cite{odic_publication_2020, gonzalez2019contemporary,mayer2018does}, revealing patterns of underrepresentation, differential citation rates, and uneven editorial review processes. 

While many studies in our sample focus on gender inequalities within a specific field, others adopt interdisciplinary approaches by examining the issue across multiple disciplines. 
In fields such as Science, Technology, Engineering, and Mathematics (STEM) \cite{king_pandemic_2021,huang_historical_2020,holman2018gender} and Science and Technology Studies (STS)\cite{abramo_gender_2009}, scholars have employed mixed methods approaches for AND by combining quantitative citation metrics with qualitative assessments of institutional practices to uncover persistent gender imbalances. Cross-domain analyses further illustrate that gender bias manifests in multiple contexts, consistently affecting authorship credit, editorial decisions, and scholarly influence \cite{pilkina_gender_2022,lariviere_bibliometrics_2013,liu_gender_2023,ross_women_2022,nielsen_gender_2016,king_men_2017,squazzoni2021gender}. 
In summary, the studies investigating gender bias span multiple domains, reflecting widespread interest across academic fields. This breadth of research demonstrates that gender bias is a pervasive issue, with researchers employing varied analyses to uncover patterns of inequity in representation, research visibility, and scholarly recognition. 

\subsubsection{Geographical and temporal scope of data used in studies}
The geographical scope of the studies on gender inequality in academia varies widely, with some papers adopting a global perspective while others opt for a country-specific approach. This breadth mirrors the broader quest for a comprehensive view of this persistent issue as well as a detailed, context-specific understanding that could inform policy changes and support initiatives attracting and retaining women in academia. Studies that have adopted a global perspective (i.e. covering multiple nations) have unveiled international trends and contributed to a broad overview of the gender inequality landscape in academia through comparative analyses \cite{caplar_quantitative_2017, lariviere_bibliometrics_2013, teich_citation_2022, king_pandemic_2021, huang_historical_2020, king_men_2017, mishra_self-citation_2018, odic_publication_2020, mohammad-2020-gender, holman2018gender,gonzalez2019contemporary,mihaljevic2016effect, bendels2018gender, card2020referees, filardo2016trends, lerchenmueller2019gender, yang2022gender,vasarhelyi_gender_2021}. 


Several studies have focused on gender inequality in academia within particular countries. For instance, \citet{pilkina_gender_2022} focused on Russia, while \citet{ross_women_2022, mcdermott2018sex,carter2017gender,gayet2019comparison} analyzed the gender imbalance within the context of the United States. \citet{abramo_gender_2009} concentrated on Italy, \citet{lariviere_sex_2011} on Quebec, \citet{rao_gender_2021} on Canada, \citet{nielsen_gender_2016} on Denmark, and \citet{mayer2018does} on Germany. These country-specific studies highlight the importance of a locally contextualized understanding of gender inequality, considering the unique cultural, social, and legislative landscapes that might influence this issue. They are an essential complement to global studies, providing a richer and more detailed picture of gender inequalities in academia. 

\subsubsection{Scholarly performance indicators}
Numerous bibliometric indicators have been used to investigate gender bias. Many studies have scrutinized female and male researchers' scholarly influence and prominence, and these assessments often employ a set of bibliometric and altmetric indicators \cite{raj2016longitudinal,fulvio2021gender,thomas2019gender}. Traditional bibliometric metrics include citation counts, patterns of self-citation and citing behavior, the impact factor of the publishing journal, and the h-index. More contemporary measures, such as social media metrics, are increasingly being utilized to capture a broader academic impact, visibility, and attention that publications achieve \cite{mcdermott2018sex,carter2017gender}. We observed that the most frequently analyzed scholarly performance indicators included the number of publications, citations, self-citations, journal impact factor, h-index, and author by-line order. 

Analyzing authors' productivity, we found that the number of publications was frequently used as a proxy for productivity \cite{zhang2022gender, odic_publication_2020,abramo_gender_2009,lariviere_sex_2011,pilkina_gender_2022, bendels2018gender}. For instance, \citet{abramo_gender_2009} found that male researchers outperformed female researchers in productivity by an average output of 16.8\% more papers. Similarly, \citet{odic_publication_2020} found that male first authors achieve higher publication and citation counts even when affiliations are controlled for. Furthermore, \citet{bendels2018gender} found that women publish fewer articles than men, hold fewer prestigious authorship positions and that articles with female key (first or last) authors are cited less frequently than those with male key authors, with these disparities being most pronounced in high-impact journals and highly collaborative articles.

To measure the impact of research paper, we observed that papers used numerous metrics such as the number of citations \cite{kozlowski2022intersectional, kong2022influence,card2020referees, mohammad-2020-gender, nielsen_gender_2016, maliniak_gender_2013, huang_historical_2020, pinheiro_women_2022, teich_citation_2022, chatterjee_gender_2021, caplar_quantitative_2017,dworkin_extent_2020,wang_gendered_2021}, h-index \cite{carter2017gender,mcdermott2018sex}, 
self-citation \cite{mishra_self-citation_2018, king_men_2017, azoulay_self-citation_2020}, and journal impact factor \cite{mihaljevic2016effect, mayer2018does}. For the number of citations, \citet{chatterjee_gender_2021} found that articles written by women as first authors had fewer median citations than those written by men as first authors. Similarly, \citet{caplar_quantitative_2017} found that papers authored by women receive 10.4 ± 0.9\% fewer citations than would be expected if men wrote the papers with the same non-gender-specific properties. For the h-index, \citet{carter2017gender} found that the gender gap in the h-index was the largest at the full professor level and smallest at the associate professor level, where women’s h-index scores were close to those of men. Also, \citet{mihaljevic2016effect} discovered that female authors published significantly less in top journals than their male counterparts. 

Some studies \cite{ahmadia2021limited, fox2018patterns,nunkoo2020three, salerno2019male, merriman2021gender,palser2022gender, gonzalez2019contemporary,edwards2018gender, lariviere_bibliometrics_2013, king_pandemic_2021, murphy2020open} looked into the authorship positions and author by-line order to investigate gender biases. Authorship positions, such as first, middle, or last author, often carry distinct levels of recognition and prestige within academic publications. In contrast, the author by-line order refers to the sequence in which authors are listed, providing insight into collaboration dynamics and the relative contributions of each author. By analyzing these aspects, these studies aimed to uncover patterns of gender disparity in recognition, leadership roles, and collaborative influence within scholarly publishing.

\citet{lariviere_bibliometrics_2013}found that articles first-authored by men outnumber those first-authored by women by nearly two to one (1.93). \citet{filardo2016trends} observed that the representation of women among first authors of original research in high-impact general medical journals was significantly higher in 2014 than 20 years ago, but has plateaued in recent years and declined in some journals. Similarly, \citet{murphy2020open} discovered that women are more likely to be represented in high-status author positions in open science. 

\end{document}